\documentclass[sn-mathphys]{sn-jnl}

\usepackage{amssymb}
\usepackage{amsmath}
\usepackage{amsfonts}
\usepackage{graphicx}
\usepackage{subcaption}
\usepackage{mathrsfs}
\usepackage{dcolumn} 
\usepackage{bm}
\usepackage{color}
\usepackage{cancel}
\usepackage{mathbbol}
\usepackage{epsfig}
\usepackage{units}
\usepackage{esint}
\usepackage{soul} 
\usepackage{url}
\usepackage{hyperref}

\jyear{2021}

\definecolor{rot}{rgb}{0.75,0.05,0.25}
\definecolor{hellgrau}{gray}{0.5}
\definecolor{blau}{rgb}{0,0,0.7}

\newcommand{\av}[1]{\left\langle#1\right\rangle}

\def\Tr{\mbox{Tr}}

\begin{document}

\title[Spontaneous fluctuation-symmetry breaking and the Landauer principle]{Spontaneous fluctuation-symmetry breaking and the Landauer principle}

\author[1]{\fnm{Lorenzo} \sur{Buffoni}}\email{lorenzo.buffoni@unifi.it}
\author[2,1]{\fnm{Michele} \sur{Campisi}}\email{michele.campisi@nano.cnr.it}

\affil[1]{Dipartimento di Fisica e Astronomia, Universit\'a degli Studi di Firenze, I-50019 Sesto Fiorentino, Italy}
\affil[2]{NEST, Istituto Nanoscienze-CNR and Scuola Normale Superiore, I-56127 Pisa, Italy}

\abstract{We study the problem of the energetic cost of information erasure by looking at it through the lens of the Jarzynski equality. We observe that the Landauer bound, $\langle W \rangle \geq kT \ln 2$, on average dissipated work $\langle W \rangle$ associated to an erasure process, literally emerges from the underlying second law bound as formulated by Kelvin, $\langle W \rangle \geq 0$, as consequence of a spontaneous breaking of the Crooks-Tasaki fluctuation-symmetry, that accompanies logical irreversibility. We illustrate and corroborate this insight with numerical simulations of the process of information erasure performed on a 2D Ising ferromagnet.}

\maketitle

\section{Introduction}
Since the seminal work of Landauer \cite{Landauer61IBMRD5}, the question regarding the energetic cost of irreversible logical operations, e.g., the erasure of a bit of memory, has become a classic in the thermodynamics of computation \cite{Bennett82IJTP21}.

According to Landauer, in order to erase a bit of information stored in a register being immersed in an environment at temperature $T$, an amount of energy of at least the order of $kT$ must be spent \cite{Landauer61IBMRD5}. More precisely, thermodynamic arguments suggest that at least $kT \ln 2$ of work must be invested:
\begin{align}
 W \geq k T\ln 2 \, .
 \label{eq:Landauer}
\end{align}

Here we address the issue of the energetic cost of information erasure by focussing on the implementation of the \texttt{RESET TO ONE} process in an Ising ferromagnet. Using the Ising  model, featuring  first and second order phase transition, to investigate the link between thermodynamics and information theory, was first proposed in the work of Ref. \cite{Parrondo01CHAOS11}, which clarified the thermodynamics of the Szilard engine, and illuminated the role that spontaneous symmetry breaking (SSB) plays in the process of information erasure. See also Ref. \cite{Roldan14NATPHYS10} for recent theoretical and experimental developments of that idea. 

At variance with the work of Ref. \cite{Parrondo01CHAOS11} that was carried at a time when work fluctuations were not yet in the limelight of non-equilibrium thermodynamic research, here we will look at Landauer's erasure principle  Eq. (\ref{eq:Landauer}) through the lens of the Jarzynski equality \cite{Jarzynski97PRL78}, which, for a cyclical process, such as the \texttt{RESET TO ONE}, reads:
\begin{align}
\langle{e^{-\beta W}}\rangle =1\, .
\end{align}
Notoriously, this combined with Jensen's inequality, implies the bound
\begin{align}
\langle W \rangle \geq 0\, ,
\label{eq:2ndLAW}
\end{align}
which is universally recognised to express the second law of thermodynamics in the formulation of Kelvin \cite{Allahverdyan02PHYSA305,Campisi11RMP83,Jarzynski97PRL78}. That is, one cannot extract energy with any generic cyclic transformation from a system in contact with a single thermal bath. Note that, within this picture the lower bound, $0$, is achieved in the quasi-static limit. 
The fact that in the \texttt{RESET TO ONE} process the lower bound is not simply that dictated by the second law, Eq. (\ref{eq:2ndLAW}), but is lifted by an amount $kT \ln 2$, suggests that there is an extra source of dissipation, which is associated with information erasure, that cannot be eliminated by decreasing the speed of the process. This observation adheres perfectly to Landauer's  view that 
``Computing, like all processes proceeding at a finite rate, must involve dissipation. [...] however [...] there is a minimum heat generation, independent on the rate of the process'' \cite{Landauer61IBMRD5}. 
One might wonder, then, what is the mechanism that lifts the bound from zero to $kT \ln 2$ when information is erased. Arguably, that is related to the absence of a quasi-static limit, a situation that typically occurs when some time-scale diverges and ergodicity breaks, e.g., as a consequence of spontaneous symmetry breaking \cite{Palmer82AdP31}, as Ref. \cite{Parrondo01CHAOS11} and the recent experiment of Ref. \cite{Gavrilov16EPL114} suggest.

With this work we investigate this insight further and put forward the idea, corroborated by extensive numerical simulations, that the Landauer bound (i.e. $kT \ln 2$, Eq.(\ref{eq:Landauer})) literally emerges from the underlying second law bound (i.e., $0$, Eq.(\ref{eq:2ndLAW})), as a consequence of a mechanism of spontaneous breaking of the symmetry that is at the basis of Jarzynski equality, namely the fluctuation symmetry, a.k.a., the work fluctuation relation \cite{Crooks99PRE60,Tasaki00arXivb,Campisi11RMP83}.

\section{The Conjecture}
The issue of the emergence of Landauer's bound, Eq. (\ref{eq:Landauer}), from the underlying second law bound, Eq. (\ref{eq:2ndLAW}), is similar to a classic issue in statistical physics, namely how a net magnetisation $\langle M_z \rangle$ is possible in an Ising ferromagnet at thermal equilibrium when its free energy is invariant under the reversal of the magnetic field \cite{GoldenfeldBook}. It is now clear that while for any finite sample size the net magnetisation is null at zero applied external field, it presents a discontinuity at zero field in the thermodynamic limit. Namely the limits of vanishing field and infinite size of the sample do not commute:
\begin{align}
0 = \lim_{N\rightarrow \infty}\lim_{h\rightarrow 0^\pm } \langle M_z \rangle \neq \lim_{h\rightarrow 0^\pm}\lim_{N\rightarrow  \infty } \langle M_z \rangle = \pm M_0.
\label{eq:spontansous-m}
\end{align}
This is the mechanism at the basis of spontaneous symmetry breaking. 

A similar phenomenon is at the basis of the process of erasure. Let us take a close look at how erasure is realised in practice in a uni-axial Ising ferromagnet. Let's say positive macroscopic magnetisation encodes the \texttt{ONE} state and negative macroscopic magnetisation encodes the \texttt{ZERO} state. Below the Curie temperature, the \texttt{RESET TO ONE} operation is implemented in practice by simply switching on an external magnetic field along the positive magnetisation axis, and then switching it off, thus implementing a cyclical and time-symmetric protocol. 
According to this procedure, regardless of the initial magnetisation of our sample, it will in the overwhelming majority of repetitions of the protocol end up in the \texttt{ONE} state. The crucial point to notice is that, in the thermodynamic limit, it will \emph{always} end up there, giving origin to a perfect erasure of information \cite{GoldenfeldBook}.

When that happens, the Jarzynski equality however would break.
To understand that, let us recall a well known fact that often makes the applicability of Jarzynski equality a difficult task: The statistical average of the exponential $e^{-\beta W}$ is dominated by rare events for which $W<0$. In fact, as established by Jarzynski \cite{Jarzynski06PRE73}, the number of realisations $\mathcal N$ necessary to efficiently sample such dominant rare events goes, for a cyclic and time-symmetric process, like $e^{\beta \langle{W}\rangle}$. So in case of $\langle{W}\rangle$ scaling, e.g., linearly with some parameter $N$, the number of realisations that need to be sampled, scales exponentially in $N$, $\mathcal N \propto e^{\beta w N}$, where $w$ is a constant prefactor. In the $N\rightarrow \infty$ limit, and at finite $\beta$, no matter how large is your statistical sample, you are going  to miss the dominant rare events. As a result, the Jarzynski equality undergoes a spontaneous symmetry breaking. In formulae:
\begin{align}
1= \lim_{N\rightarrow \infty} \lim_{\mathcal N\rightarrow \infty} \langle e^{-\beta W} \rangle_{N, \mathcal N}  
\neq
\lim_{\mathcal N \rightarrow \infty} \lim_{N\rightarrow \infty} \langle e^{-\beta W} \rangle_{N, \mathcal N} = \gamma 
\label{eq:JE-nN-Nn}
\end{align}
with some $\gamma \leq 1$, where $\av{\cdot}_{N,\mathcal N}$ denotes the empirical average obtained from a finite sample of size $\mathcal N$ for a system of size $N$. That is:  $\langle e^{-\beta W} \rangle_{N, \mathcal N}=(1/\mathcal N)\sum_{i=1}^{\mathcal N} e^{-\beta W_i}$ with $W_i$ denoting the value of work of item $i$ in the sample\footnote{Note that $\langle e^{-\beta W} \rangle_{N, \mathcal N}$ is a stochastic quantity, that well represent the measured exponential average in any experiment that necessarily is based on a finite statistical sample of size $\mathcal N$.}.
Eq. (\ref{eq:JE-nN-Nn}) has to be understood in the following way. If you have an ideally infinitely large statistical sample, $\mathcal N = \infty$, no matter how large is $N$ you will observe $\langle e^{-\beta W} \rangle=1$; this is the first equality in Eq. (\ref{eq:JE-nN-Nn}). If you have a finite statistical sample (no matter how large), but $N=\infty$, then you will observe $\langle e^{-\beta W} \rangle=\gamma$; this is the second equality in Eq. (\ref{eq:JE-nN-Nn}).
In this latter case, we say that the fluctuation symmetry (the Tasaki-Crooks relation \cite{Crooks99PRE60,Tasaki00arXivb})
which is at the basis of the Jarzynski equality undergoes a spontaneous breaking. As will be detailed below, for an Ising ferromagnet below the critical temperature, the parameter $N$ that drives the symmetry breaking is the size $N$ of the system, but it can be some other quantity in different physical implementations of a memory.

According to the fluctuation symmetry, each dynamical trajectory $y$ has a time-reversal conjugate $\tilde y$, and their respective probabilities are linked (for cyclic and time-reversal symmetric forcing protocols) by the relation
\begin{align}
\frac{p(y )}{p(\tilde y)}= e^{\beta W(y)}
\end{align}
with $W(y)$ the work associated to the trajectory $y$. For an erasure process, the conjugates of trajectories $\tilde y$ associated to large (order $N$ or larger) work $W$, have some finite probability at finite size, but
are \emph{de facto} excluded from the statistics (have zero probability) in the t$N \rightarrow \infty$ limit.  The phenomenon whereby some trajectories do not have their mirror image companion is referred to in the literature as ``absolute irreversibility'' \cite{Hoang16SCIREP6,Murashita14PRE90}.

Using Jensen's inequality with Eq. (\ref{eq:JE-nN-Nn}) one gets two distinct bounds to the work:
\begin{align}
\lim_{N\rightarrow \infty} \lim_{\mathcal N\rightarrow \infty} \langle W \rangle_{N, \mathcal N} &\geq 0, \label{eq:landauer-no-SSB}\\
\lim_{\mathcal N \rightarrow \infty} \lim_{N\rightarrow \infty} \langle W \rangle_{N, \mathcal N} &\geq - k T\ln \gamma \geq 0 .\label{eq:landauer-SSB}
\end{align}
The top line expresses the second law, the bottom line expresses Landauer's principle. 
The former has a fundamental status and universal validity, in fact, since $\gamma \leq 1$ it is always true that $\langle W \rangle \geq 0$. 
We remark that there is actually no issue in regard to commutation of limits for the quantity $ \langle W \rangle_{N, \mathcal N} $, and in fact the two equations above can be conveniently condensed into a single inequality
\begin{align}
 \langle W \rangle &\geq - kT \ln \gamma ,
\end{align}
where it is understood that $\gamma=1$ in absence of information erasure and SSB of the fluctuation symmetry, and gets a lower value $0<\gamma<1$ instead.

%


This situation is the non-equilibrium analogue of the observed spontaneous magnetisation in an Ising ferromagnet at equilibrium. For the latter, the time scale associated to a sign reversal goes like
\begin{align}
\tau_0 \sim e^{N^{(d-1)/d}},
\end{align}
with $d$ the spatial dimensions (see Ref. \cite{GoldenfeldBook}, Sec. 2.10). So, practically, for $d>1$ and sufficiently large systems, one is never going to see a reversal, and conclude on empirical basis, that on average, the magnetisation is not null. Similarly, in the erasure process, one can never see the rare events and conclude on empirical ground that the second law bound should be lifted from zero to $-kT \ln \gamma$.
\begin{figure}[]%
    \centering
    \includegraphics[width=0.6\linewidth]{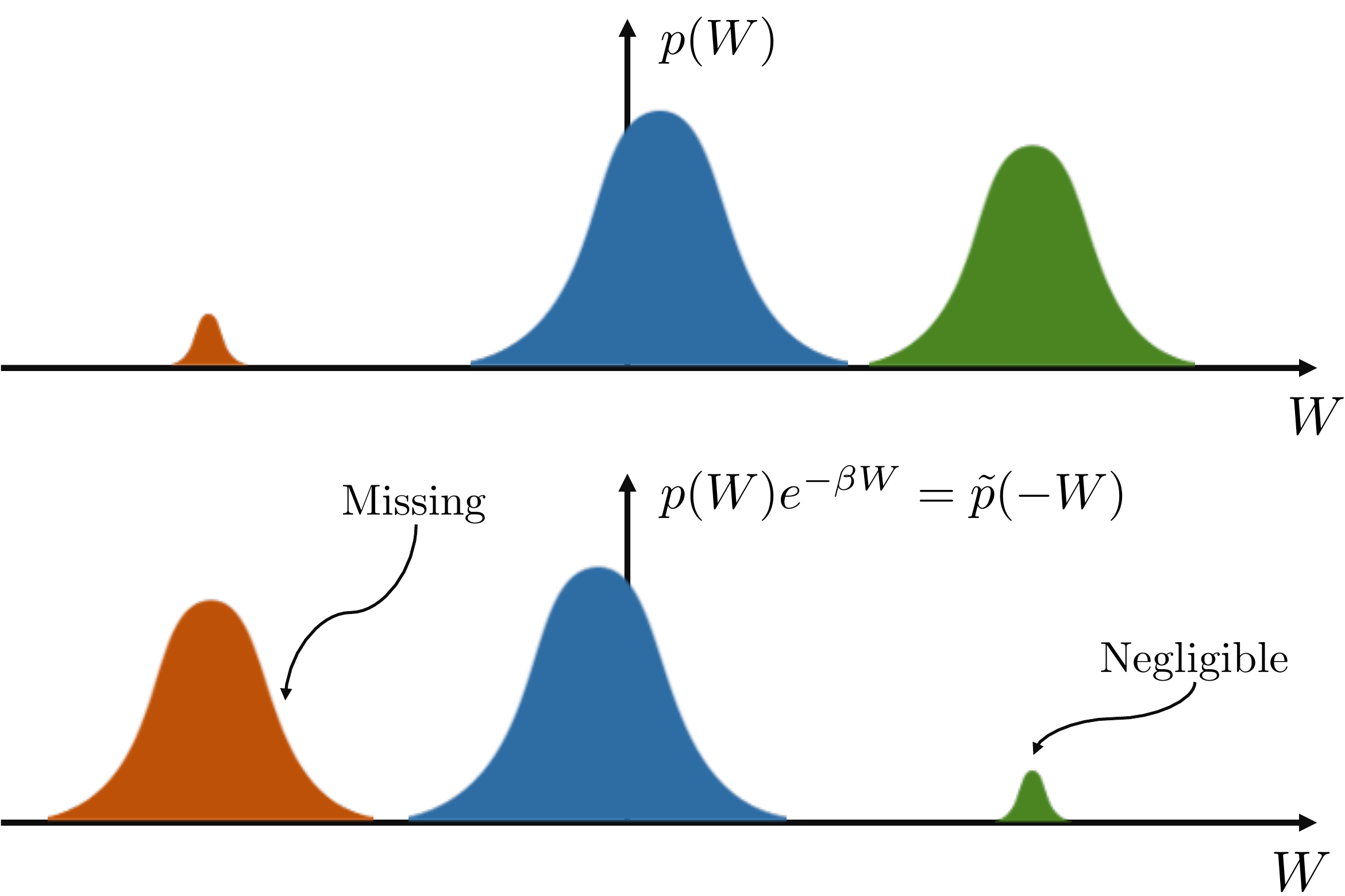} 
    \caption{Mechanism of spontaneous breaking of the fluctuation symmetry. Top: Sketch of a work pdf $p(W)$. Bottom: Sketch of the time-reversed companion of $p(W)$, namely $\tilde{p}(-W)=p(W)e^{-\beta W}$. The exponential average $\langle e^{-\beta W} \rangle=\int dW e^{-\beta W}p(W)$ is contributed about half and half by the central and left peaks. As we argue, in a real experiment, the left peak is missing due to lack of statistics. Accordingly the \emph{empirical} value is contributed by the central peak only, $\langle e^{-\beta W} \rangle \simeq 1/2$, while the true ideal value is $\langle e^{-\beta W} \rangle= 1$. The colours refer to $\mathcal S_+ \rightarrow \mathcal S_-$ transitions (orange), $\mathcal S_- \rightarrow \mathcal S_+$ transitions (green) and no-transitions (blue).}
    \label{fig:work-pdf}%
\end{figure}

\begin{figure}[t]
    \centering
    \begin{subfigure}{0.48\textwidth}
		\centering
		 \includegraphics[width=\linewidth]{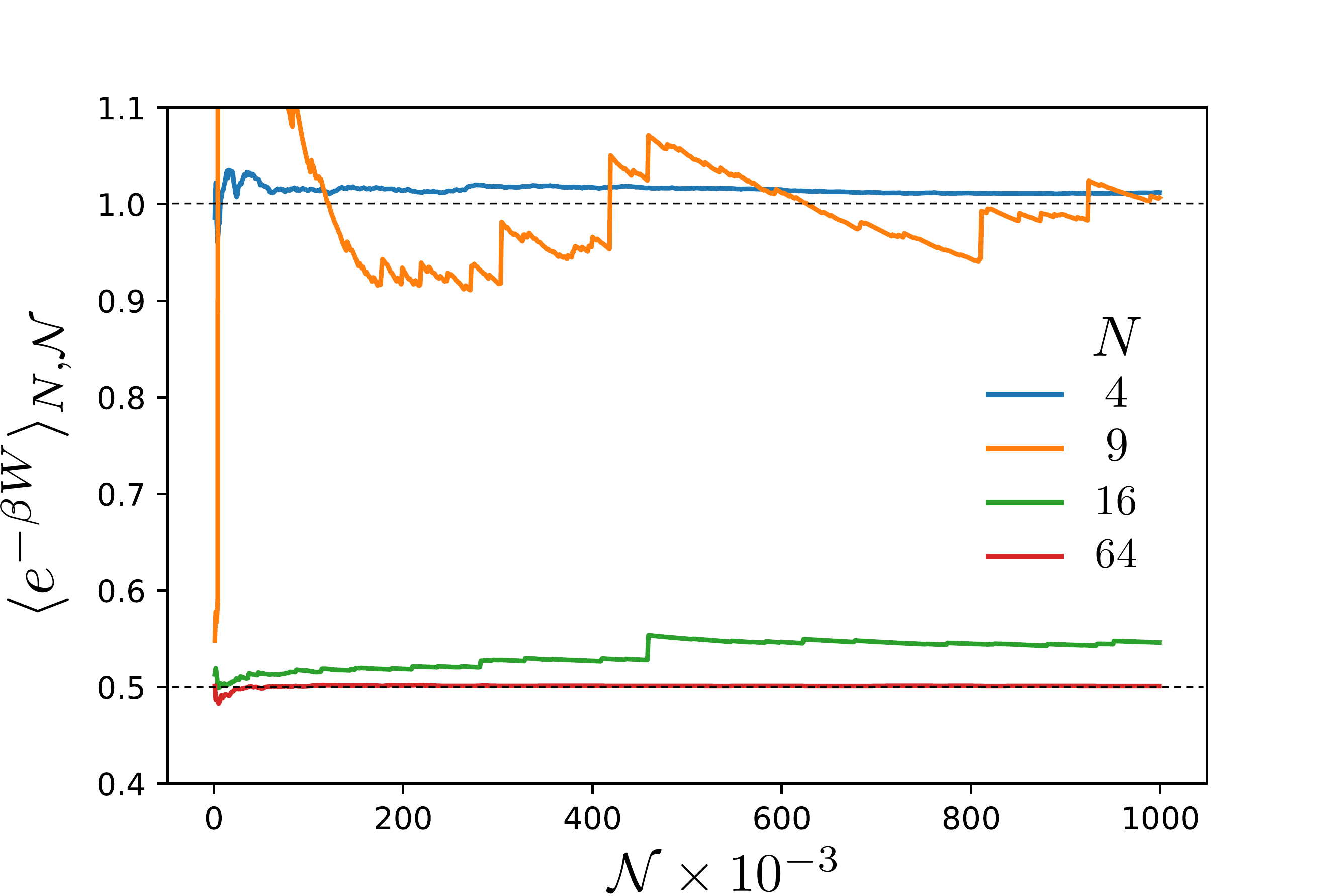}
		\caption{}
    \end{subfigure}
   \begin{subfigure}{0.49\textwidth}
   		\centering
    		\includegraphics[width=\linewidth]{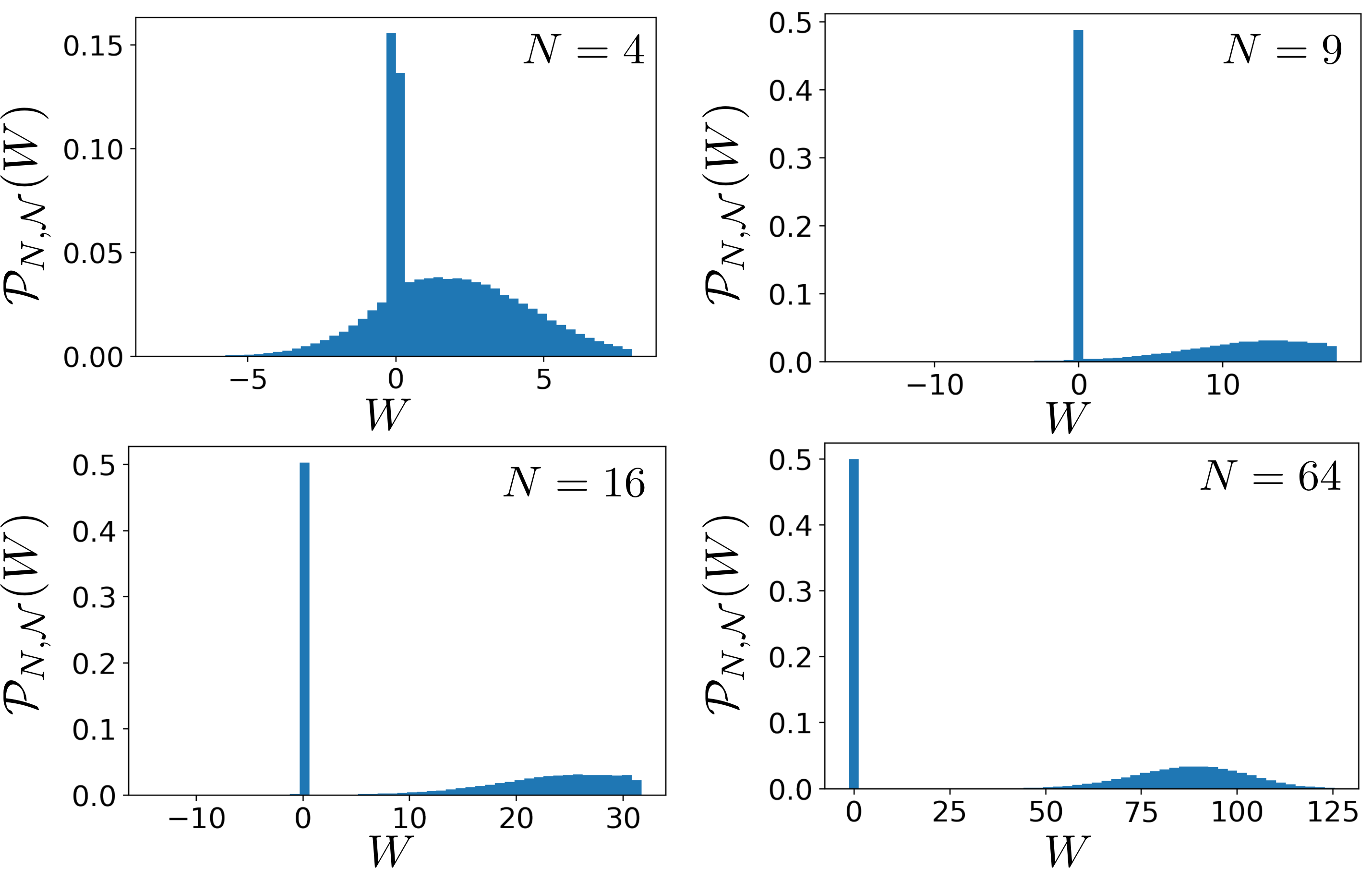}
		\caption{}
    \end{subfigure}
    \caption{Panel (a): Average exponentiated work $\langle e^{- \beta W(y)}\rangle_{N, \mathcal N}$  (solid lines) as a function of $\mathcal N$ for a reset protocol on a square lattice of different sizes $N$. The top dashed line represent the value $1$ while the bottom dashed line the value $1/2$. Panel (b): work distributions $\mathcal P_{N,\mathcal N}(W)$ at different sizes $N$ for $\mathcal N =10^6$, $h_0=1 ,\tau = 200$, $T=1.5$. }
    \label{fig:dist}
\end{figure}

For a memory featuring two symmetric states, $\gamma$ would take on the value $1/2$, and the second line in Eq. (\ref{eq:landauer-SSB}) would boil down to the Landauer principle in the form of Eq. (\ref{eq:Landauer}). 
To see that, consider the case of a uni-axial Ising ferromagnet. The configuration space can be split in two sectors $\mathcal S_{\pm}$ pertaining to positive or negative magnetisation, respectively. Hence there are 4 main families of events that can take place during the cyclic process of switching the filed on and off:
a) $\mathcal S_+ \rightarrow \mathcal S_+$, b) $\mathcal S_+ \rightarrow \mathcal S_-$, c) $\mathcal S_- \rightarrow \mathcal S_+$, d) $\mathcal S_- \rightarrow \mathcal S_-$. 
Let us consider the case in which we are below the critical temperature. Events of the type $\mathcal S_+ \rightarrow \mathcal S_+$, apart from fluctuations, are such that the system goes along a branch of the hysteresis loop and then traces it back. These are reversible processes, with ideally small dissipation. They are very frequent processes, occurring almost every time the system already starts in $\mathcal S_+$, hence with a probability $ \lesssim 1/2$.
Events of the type $\mathcal S_+ \rightarrow \mathcal S_-$, roughly correspond to traversing a hysteresis branch in opposite direction, which are associated with an extensive gain in energy, i.e., a large negative dissipation. They are extremely rare, and their relative frequency is expected  to vanish in the thermodynamic limit.
Events of the type $\mathcal S_- \rightarrow \mathcal S_+$ correspond to traversing half of a hysteresis loop, they are very typical and frequent, and are associated to a positive dissipation. They occur almost every time the system starts in $\mathcal S_-$, hence with a probability $\lesssim 1/2$.
Events of the type  $\mathcal S_- \rightarrow \mathcal S_-$ correspond to very small dissipation, and are very infrequent.
Thus, roughly, one might expect tri-modal work PDF as the one sketched in Figure \ref{fig:work-pdf}, with a peak centered around a value  $W\gtrsim0$, one peak around a positive extensive value $W\simeq N\omega_0$, and a peak at $W\simeq - N\omega_0$. The central and right peak contain basically all the probability, and approximately share it equally, that is they both enclose an area of about $1/2$. Due to Crooks relation, the left peak is exponentially smaller than the right one.
Now when calculating $\langle e^{-\beta W} \rangle$, the left peak counts exponentially more that the right peak. The result is that, since the events in the left peak are so rare you are missing them all, and the events in the right peak do not count, only the events in the central peak count in practice. The integral $\langle e^{-\beta W} \rangle$, in real experiments on a large but finite sample, amounts then to the probability of central peak only which is roughly 1/2, hence $\gamma \simeq 1/2$. Above the critical temperature it is expected that only the central peak is there, or other peaks exist but are not too far separated by each other. Accordingly, there is no missing statistics and one should get $\langle e^{-\beta W} \rangle =1$. 

We stress that the above argument on the work PDF is not meant to be mathematically rigorous, but only to illustrate the main idea behind the process of symmetry breaking. In the following we report the results of numerical simulations on the 2D Ising model, display the actual form of the work PDF (see Figure \ref{fig:dist}b) and give a precise estimate of how hard it is to sample the rare events from our statistics.

It is important to remark that, albeit in the case of the Ising ferromagnet the size $N$ of the system is the parameter that drives the fluctuation-symmetry breaking, the fluctuation symmetry breakdown may be driven by other parameters depending on the specific physical scenario considered for erasure.  The thermodynamic limit is not essential for our analysis.
For example, with reference to the erasure of information stored in a single particle in a double-well potential studied e.g., in Refs. \cite{Berut12Nature483,Jun14PRL113,Berut15JSTAT,Hong16SCIADV2,Wimsatt21PRR3}, the symmetry-breaking parameter would be the height of the energy barrier $E$ separating the two logical states.\footnote{In the Ising model, the two logical states are effectively separated by an energy barrier of order $N^{(d-1)/d}$, see Sec. 2.10 of Ref. \cite{GoldenfeldBook}, thus increasing $N$ increases the energy barrier for $d \geq 2$.} The difficulties in sampling rare trajectories in an erasure process and so verify the Jarzynski equality in that case were in fact well illustrated by the experiments reported in Refs. \cite{Berut15JSTAT,Berut13EPL103}, we shall comment further on this in Sec. \ref{subsec:relation}.

\section{The Numerics}

In the following we corroborate our argument with numerical experiments.
To this end we simulated the dynamics of a 2D Ising ferromagnet on a square lattice of side length $L$, evolving according to Glauber dynamics \cite{Glauber63JMP4}.
The Hamiltonian reads:
\begin{align}
H(t) =  h(t) \sum_i \sigma_i^z - J \sum_{\langle i,j \rangle} \sigma_i^z \sigma_j^z.
\end{align}

In our simulations the initial state of the system is randomly sampled from a Gibbs thermal equilibrium at a temperature $T$ and the \texttt{RESET TO ONE} protocol is performed by linearly ramping up $h(t)$ to some positive value $h_0$ in a time $\tau/2$ and linearly ramping down to zero in the same time $\tau/2$, so that our protocol is cyclic and time-symmetric.
Each randomly sampled initial state of the system is evolved according to Glauber dynamics with temperature $T$, so as to produce a trajectory $y= \{S_i(t)\}_{i=1\dots N}^{t\in[0,\tau]}$, with $N=L^2$. The dynamics of the aforementioned model was generated using a Markov Chain Monte-Carlo approach which we parallelized using the python package numba \cite{lam2015numba} to increase our sampling capabilities (the code we used to run our numerical experiments is publicly available and can be found at this \href{link}{https://github.com/Buffoni/landauer_parallel}.)

For each generated trajectory $y$ we record the net magnetisation $M(t)=\sum_i S_i(t)$ at each time, during the evolution and use the formula
\begin{align}
 W(y) = \int dt \dot{{h}}(t)  {M}(t) 
\end{align}
to calculate the according work. By repeating the erasure protocol $\mathcal N$ times we construct the statistical ensemble $\mathcal P_{N,\mathcal N}(y)$, over which we evaluate average values of trajectory dependent quantities $O(y)$:
\begin{align}
\langle O(y)\rangle_{N,\mathcal N} = \sum_y O(y)  \mathcal P_{N,\mathcal N}(y).
\end{align}

In Figure \ref{fig:dist}, we report the quantity $\langle e^{- \beta W(y)}\rangle_{N,\mathcal N}$ as a function of $\mathcal N$ for a reset protocol on a square lattice of sizes $N=[4,9,16,64]$, here $h_0= 1$ and $\tau = 200$.
Note how for the $N=4,9$ spin system, the asymptotic value $1$ is quickly approached as the sample size is increased while  picture changes drastically for $N=16$, i.e.,  by increasing the side of the lattice of a single unit. Now the average of  $\langle e^{- \beta W(y)}\rangle_{N,\mathcal N}$ remains far below the value $1$, and close to $1/2$ for the same values of $\mathcal N$. Note how the running average slowly increases with $\mathcal N$. This reflects the fact that as the sample is increased, more rare events were sampled. These are well visible as jumps in the plotted curve. Their relative number remained however greatly insufficient to guarantee convergence to the ideal value $1$. This could be achieved only with sample sizes of orders of magnitude larger than the maximal value $\mathcal N= 10^6$ employed here. In the case $N=64$ the curve $\langle e^{- \beta W(y)}\rangle_{N,\mathcal N}$ does not significantly depart from the value $1/2$. In the thermodynamic limit, there will be no departure at all. This rich phenomenology is compactly summarised in Eq. (\ref{eq:JE-nN-Nn}).
\begin{figure}[]
    \centering
    \includegraphics[width=0.65 \linewidth]{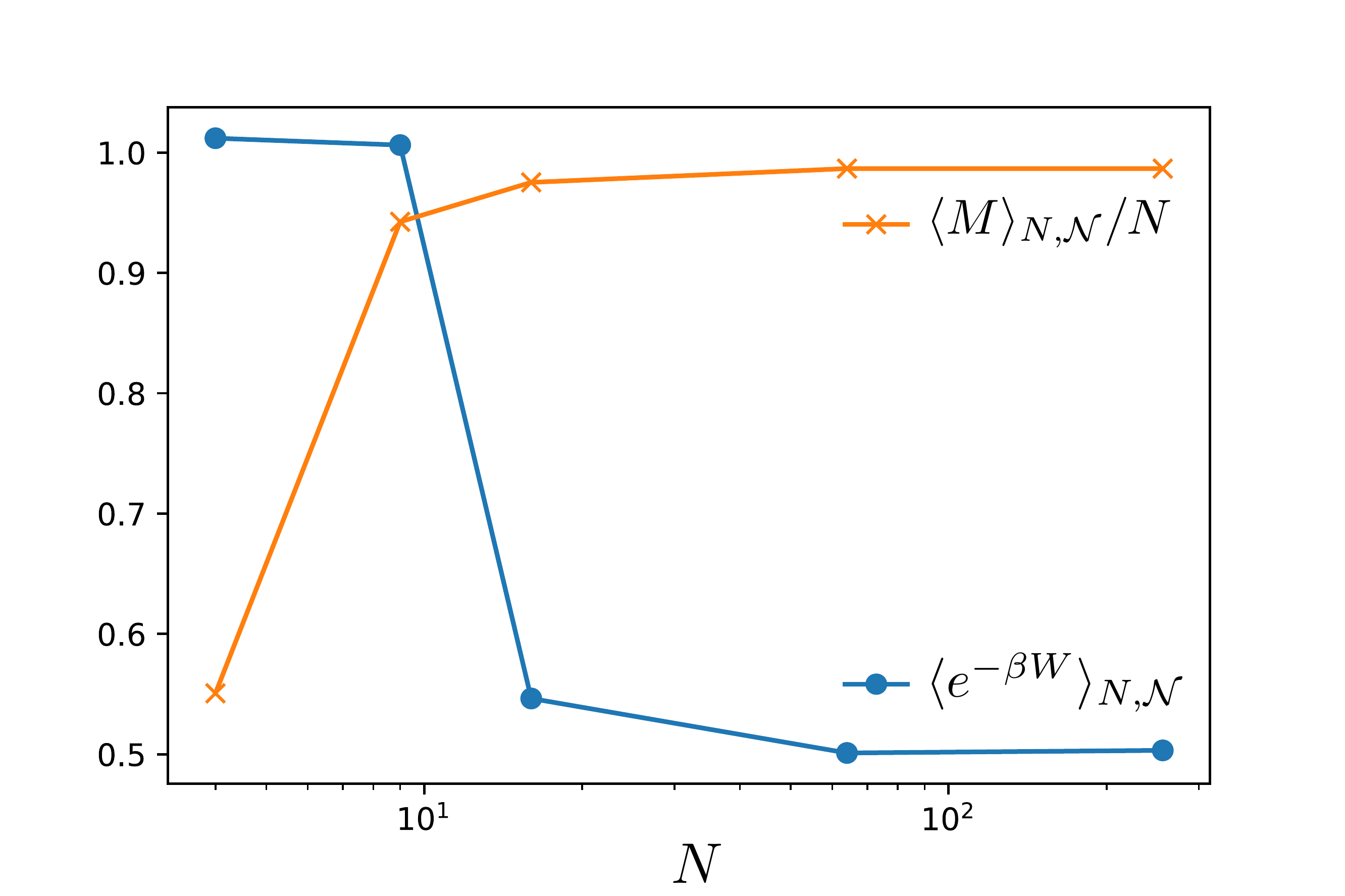}
    \caption{ Average exponentiated work $\langle e^{- \beta W(y)}\rangle_{N,\mathcal N}$ (blue dots) and magnetisation density $\langle M\rangle_{N,\mathcal N}/N$ (orange crosses), as a function of system size $N$ for a sample of size $\mathcal N=10^6$ realisations of the \texttt{RESET TO ONE} protocol with $h_0=1, \tau=200 $, $T=1.5$.}
    \label{fig:steps}
\end{figure}

The right panels of Figure \ref{fig:dist} report the plots of the work statistics $\mathcal P_{N,\mathcal N}(W)$ for $N=[4,9,16,64]$ and $\mathcal N=10^6$. Note that besides a central and a right peak, a negative tail of the distribution is visible only in the $N=4$. For $N\geq 9$, the histograms still display a central peak and a right peak, which moves farther away from the origin of the $W$, axis, while no negative tail is visible: the sampling of rare events was greatly insufficient in these cases.

\begin{figure}[t]
    \centering
        \begin{subfigure}{0.49\textwidth}
		\centering
		 \includegraphics[width=\linewidth]{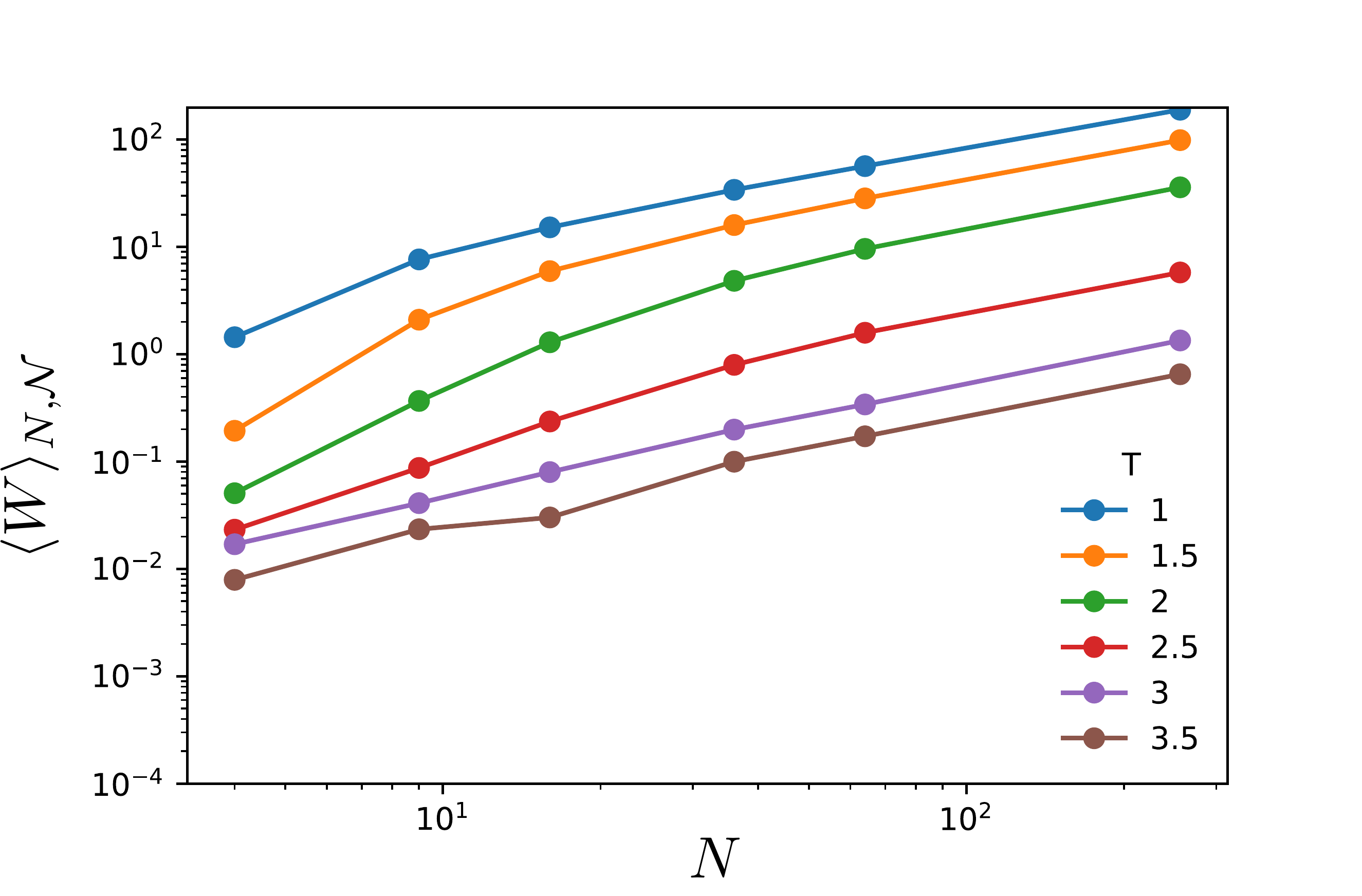}
		\caption{}
    \end{subfigure}
   \begin{subfigure}{0.49\textwidth}
   		\centering
    		\includegraphics[width=\linewidth]{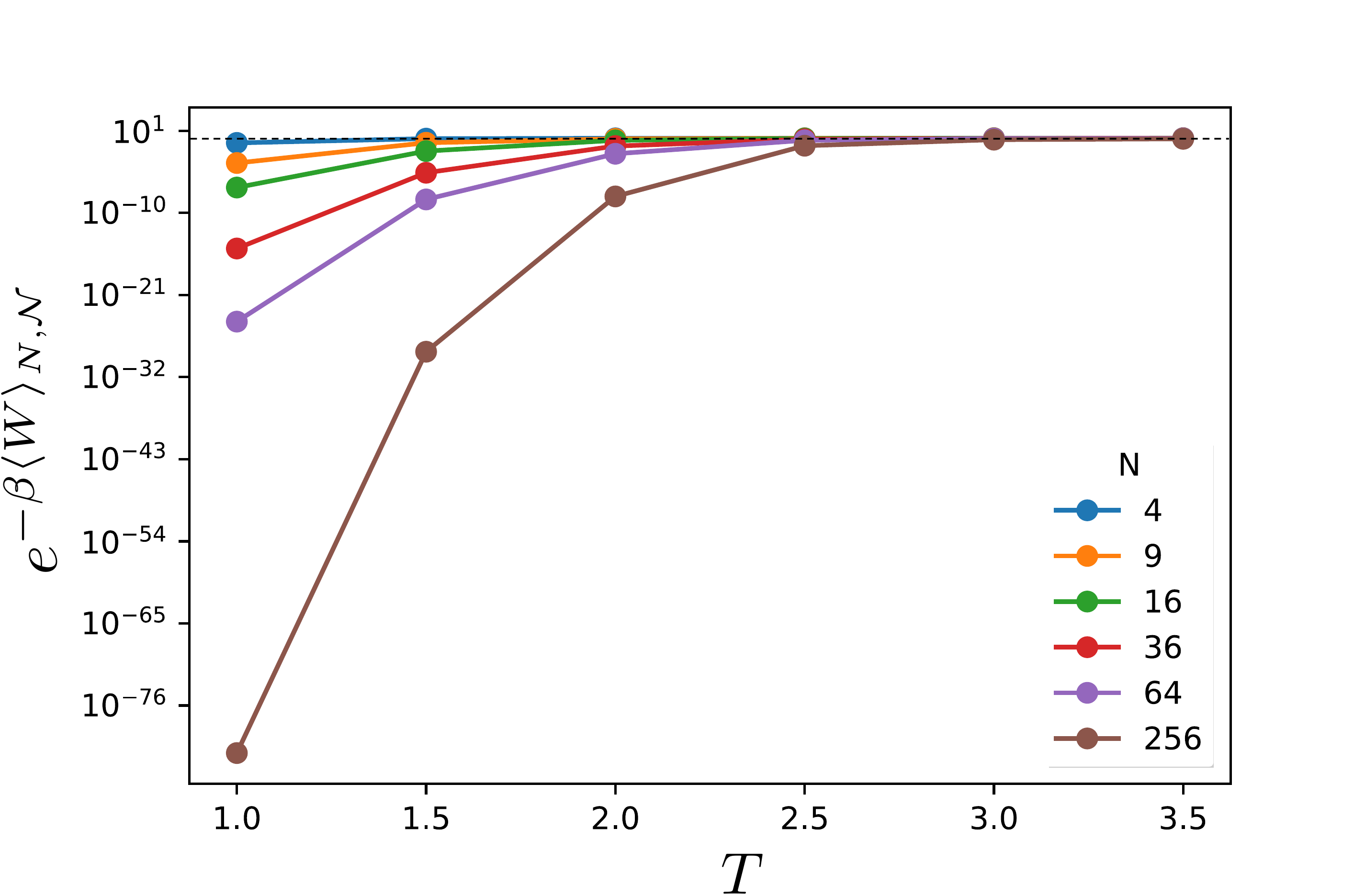}
		\caption{}
    \end{subfigure}
    \caption{Panel (a): average work $\langle W \rangle_{N,\mathcal N}$ as a function of $N$  for a fixed sample $\mathcal N$ and various temperatures, in log-log scale. Panel (b): the probability  $e^{-\beta \langle W \rangle_{N,\mathcal N}}$ of rare events, as a function of temperature $T$, for various lattice sizes $N$ at fixed $\mathcal  N =10^4$, $h_0=1 , \tau =200$. The dashed line represents probability one for comparison.}
    \label{fig:fluctuations}
\end{figure}

Figure \ref{fig:steps} shows, for a sample of size $\mathcal N = 10^6$, both $\langle e^{- \beta W(y)}\rangle_{N,\mathcal N}$ and the magnetisation density $\langle M\rangle_{N,\mathcal N}/N$ as a function of system size $N$, with $N$ up to $256$, at $T = 1.5<T_C $ (we recall that the critical temperature is $T_C\sim 2.27$). Note how, as the system size increases, the erasure protocol becomes more and more successful, i.e., the quantity $\langle M\rangle_{N,\mathcal N}/N$ approaches the value $1$, while, accordingly, the observed value of $\langle e^{- \beta W(y)}\rangle_{N,\mathcal N}$ approaches $1/2$.

Figure \ref{fig:fluctuations}, left panel, shows plots of the average work $\langle W \rangle_{N,\mathcal N}$, as a function of $N$, for fixed $\mathcal N = 10^4$ and various temperatures. All plots, in log-log scale evidence an approximately linear increase with $N$, $\langle W \rangle_{N,\mathcal N} \simeq \omega_0 N$ with coefficients $\omega_0$ that quickly vanish as the temperature is lowered below the Curie temperature $T_C \simeq 2.27$.  
Figure \ref{fig:fluctuations}, left panel, shows plots of the quantity $\mathcal P \propto e^{-\beta \langle W \rangle_{N,\mathcal N}}$, which is a rough estimate of the probability of observing a rare fluctuation \cite{Jarzynski06PRE73}. Note how far above the Curie temperature $T_C\sim 2.27 $, the fluctuation probability is always close to one, while as the temperature is decreased well below the Curie point, the decrease with $N$ becomes drastic. As discussed above the aftermath of this fact is that an exponentially large sample is needed in order to effectively sample the rare events, which, accordingly, get completely excluded from the statistics in the thermodynamic limit.

\begin{figure}[]
    \centering
    \begin{subfigure}{0.49\textwidth}
		\centering
		 \includegraphics[width=\linewidth]{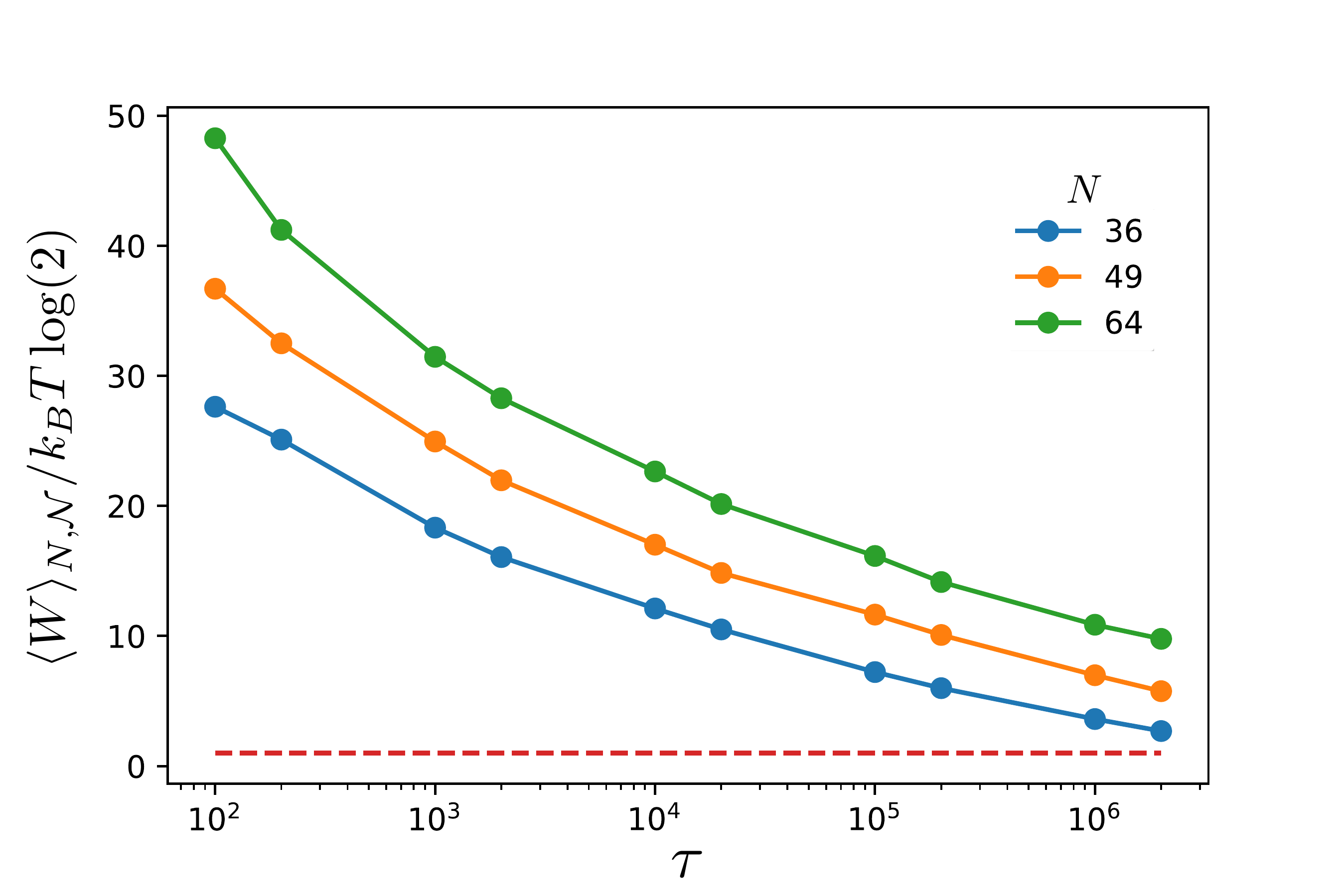}
		\caption{}
    \end{subfigure}
   \begin{subfigure}{0.49\textwidth}
   		\centering
    		\includegraphics[width=\linewidth]{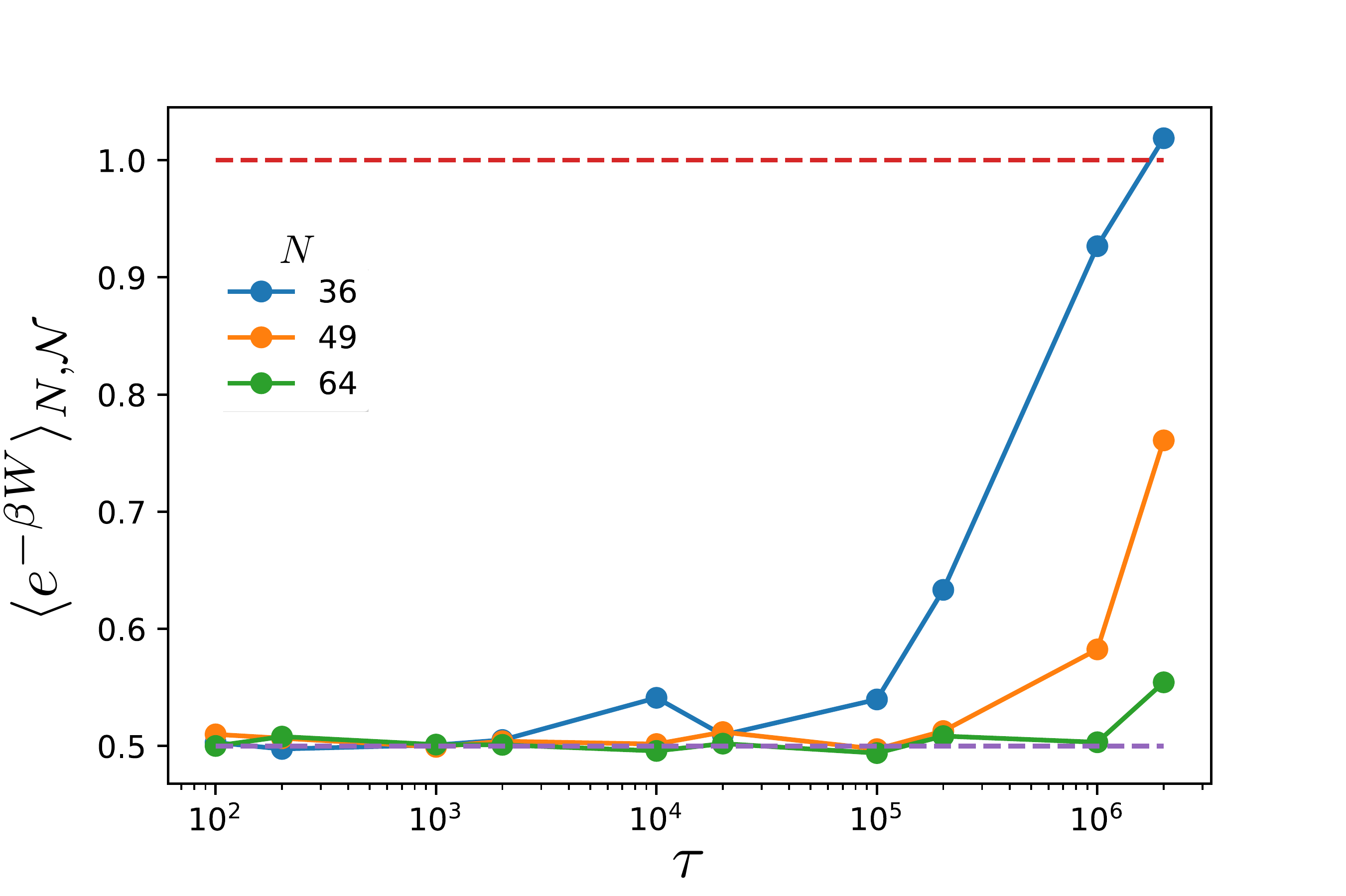}
		\caption{}
    \end{subfigure}
    \caption{Panel (a): average work $\langle W \rangle_{N,\mathcal N}/ k_B T \ln 2$ as a function of the total protocol time $\tau$ (on a log scale) and for different sizes $N$. The dashed line indicates the Landauer bound $\langle W \rangle_{N,\mathcal N}/ k_B T \ln2 = 1$ . Panel (b): average exponentiated work $\langle e^{- \beta W(y)}\rangle_{N,\mathcal N}$ as a function of  protocol duration $\tau$ (on a log scale) for various system sizes $N$. The two dashed lines indicate the values $1$ and $1/2$. For both panels we have $\mathcal N=10^4$, $h_0=1$, $T=1.5$.}
    \label{fig:adiab}
\end{figure}

Figure \ref{fig:adiab}, left panel, shows $\langle W \rangle_{N,\mathcal N}/ k_B T \ln 2$ as a function of the duration $ \tau$ of the erasure protocol, for various system sizes $N=[36,49,64]$ and fixed $\mathcal N = 10^4 $, $h_0=1$ and $T=1.5$.
We note that, as expected, the average work decreases with increasing duration. Figure \ref{fig:adiab}, right panel, shows the according average exponentiated work $\langle e^{- \beta W(y)}\rangle_{N,\mathcal N}$. The plots evidence  that in the limit of large $\tau$ the quantity departs from the value $1/2$ and approaches $1$, regardless of $N$. However, the departure from
$1/2$, occurs at a take off time that grows with $N$. This can be translated by saying that in the large $N$ limit, the quantity $\langle e^{- \beta W(y)}\rangle_{N,\mathcal N}$ remains at $1/2$, ragardless of the value of $\tau$. This suggests that the quantity $\langle W \rangle_{N,\mathcal N}$ would tend to $kT\ln 2$ in the limit of large $N$, followed by the large $\tau$ limit, and finally the large $\mathcal N$ limit.

\section{Discussion}

\subsection{Relation to previous works}
\label{subsec:relation}
\label{subsection}
The results presented above are closely related to those obtained in various previous works, reporting on experiments of erasure of information carried in double well systems \cite{Berut15JSTAT,Berut13EPL103,Gavrilov16EPL114,Wimsatt21PRR3}. Specifically, in the work of Ref. \cite{Berut13EPL103}, thanks to a coarse grained version of the fluctuation relation \cite{Vaikuntanathan09EPL87,Wimsatt21PRR3}, the averaged exponentiated work is expressed in terms of the probability, $P_S$, of success of the reset to zero process as:
\begin{align}
\langle e^{-\beta W}\rangle = P_S \langle e^{-\beta W}\rangle_{\rightarrow 0} + (1-P_S)\langle e^{-\beta W}\rangle_{\rightarrow 1} =1
\quad (0 < P_S <1)
\label{eq:Berut1}
\end{align}
where  and the symbol $\langle \cdot \rangle_{\rightarrow X}$
denotes the average conditioned on the outcome of the process being the logical state $X$, with $X=0,1$. Note that, since Eq. (\ref{eq:Berut1}) holds for any $0<P_S<1$, it holds as well in the limit $P_S\rightarrow 1$.

The authors of \cite{Berut13EPL103} also observe that, in the limit $P_S\rightarrow 1$, it is
$
\langle e^{-\beta W}\rangle_{\rightarrow 0} = 1/2.
$
Noting that, by definition, at $P_S=1 $, the conditional average coincides with the full average, namely $\langle \cdot \rangle = \langle \cdot \rangle_{\rightarrow 0}$, one gets,
\begin{align}
\langle e^{-\beta W}\rangle = 1/2 \quad (P_S=1).
\label{eq:Berut2}
\end{align}
That is, the limit of perfect erasure is a singular limit, a fact that was not noted in Ref. \cite{Berut13EPL103}. 
Here we have spelled this fact out in a most explicit manner, and expressed it in terms of lack of commutation of limits.
To make the connection more clear, perfect erasure occurs in the particle in the double well case when the height of the energy barrier that separates the two logical states diverges. The height of the barrier is accordingly the parameter that drives the fluctuation symmetry breaking in that case. If this limit is taken first, that is, if one is exactly at $P_S=1$, Eq. (\ref{eq:Berut2}) applies no matter how large is the statistical sample one is using.
Notably the experimental data collected in Ref. \cite{Berut13EPL103} adhere to the prediction of Eq.  (\ref{eq:Berut2}), rather than Eq. (\ref{eq:Berut1}) in agreement and analogy with our discussion above of the 2D Ising ferromagnet. We remark that, as the authors of Ref.  \cite{Berut13EPL103} first noted, it is Eq. (\ref{eq:Berut2}) that allows to obtain the Landauer bound, rather than the ideal Jarzynski equality (\ref{eq:Berut1}), via application of Jensen's inequality.\footnote{Our Eqs. (\ref{eq:landauer-no-SSB},\ref{eq:landauer-SSB}) express the same idea in a different language and notation.} 
Based on this observation, one might argue that the Landauer bound is a consequence of the breakdown of the Jarzynski equality, rather than a consequence of its validity.

\subsection{Landauer principle is not the second law of thermodynamics}

It is important to remark that the Landauer principle may also be obtained from an inequality first derived in Ref. \cite{Schloegl66ZP191}, (see also Refs. \cite{Kawai07PRL98,Vaikuntanathan09EPL87,Deffner10PRL105,Esposito11EPL95}) reading, for a system weakly coupled to a thermal bath and initially at thermal equilibrium:
\begin{align}
W - \Delta F \geq T D[\rho_t \vert \vert \rho_t^\text{eq}]
\label{eq:schloegl}
\end{align}
where $\Delta F= F(t)-F(0)$, with $F(t) = -\beta^{-1}\Tr e^{-\beta H(t)}$ denoting the free energy, $\rho_t$ is the  distribution describing the state of the system at time $t$, $\rho_t^\text{eq} = e^{-\beta[H(t)-F(t)]}$ is the reference equilibrium distribution at time $t$, and the symbol $D[\rho \vert \vert  \sigma]=\Tr \rho (\ln \rho-\ln \sigma)$ denotes the  relative entropy (Kullback-Leibler divergence).\footnote{While we employ a quantum notation where the symbol $\rho$ denotes a density matrix, these expressions hold as well at classical level where $\rho$ denotes a phase space density or a discrete set of probabilities and $\Tr$ denotes accordingly integration over the phase space or a sum over the events space.} Note that, since both $T$ and the relative entropy are non-negative quantities, Eq. (\ref{eq:schloegl}) implies the second law of thermodynamics:
\begin{align}
W \geq \Delta F\, .
\label{eq:W>DF}
\end{align}
Accordingly, Eq.  (\ref{eq:schloegl})  contains more information than that contained in the second law, Eq.  (\ref{eq:W>DF}).

As noted in  Ref. \cite{Esposito11EPL95}, introducing the non-equilibrium functional
\begin{align}
f[\rho_t] = E[\rho_t] - T s[\rho_t]
\end{align}
where $s[\rho_t]= -\Tr \rho_t \ln \rho_t$ is the Von-Neumann information, $E[\rho_t]=\Tr H(t) \rho_t$ is the energy expectation, 
and using the identity 
\begin{align}
f[\rho_t]-F(t) = T D[\rho_t \vert \vert \rho_t^\text{eq}]\, ,
\end{align}
Eq. (\ref{eq:schloegl}) may equivalently be rewritten as $ W - F(t)+F(0) \geq f[\rho_t] - F(t)$, or, for an initial equilibrium state, such that $f[\rho_0]=F(0)$, as:
\begin{align}
W \geq \Delta f . 
\label{eq:not-the-2nd-law}
\end{align}
Despite the appearance, the latter is not the second law of thermodynamics, because it presents the  \emph{non-equilibrium} quantity $f$ instead of the \emph{equilibrium}  thermodynamic potential $F$. For this reason, in Ref.  \cite{Esposito11EPL95} Eq. (\ref{eq:not-the-2nd-law}) is referred to as the ``non-equilibrium second law'', as opposed to the ``second law'', Eq. (\ref{eq:W>DF}).\footnote{We remark that both Eq. (\ref{eq:W>DF}) and (\ref{eq:not-the-2nd-law}) apply to non-equilibrium processes, hence the word ``non-equilibrium'' in the expression ``non-equilibrium second law''  of Ref.  \cite{Esposito11EPL95} refers to the quantity $f$ (as opposed to the equilibrium quantity $F$), rather than to a property of the considered processes.}
Note that, since (for an initial state of thermal equilibrium) it is $\Delta f \geq \Delta F$, then Eq. (\ref{eq:not-the-2nd-law}) presents a stricter bound than Eq. (\ref{eq:W>DF}).

In the limit of perfect erasure where the initial equilibrium statistics $\rho_0=\rho_0^\text{eq}$  gets finally squeezed into half of its initial support and achieves the according state of \emph{local} thermal equilibrium, $\rho^{\text{eq},X}$, it is $\Delta f = kT \ln 2$, and Eq. (\ref{eq:not-the-2nd-law}) reduces to Landauer's principle, Eq. (\ref{eq:Landauer}) \cite{Piechocinska00PRA61}. Note that the equilibrium free energy change is null in that case, $\Delta F =0$, while the second law, Eq. (\ref{eq:W>DF}) would read as Eq. (\ref{eq:2ndLAW}).
 
This last observation illustrates the crucial point that Eq. (\ref{eq:not-the-2nd-law}) is generally \emph{not equivalent} to the second law of thermodynamics, Eq. (\ref{eq:W>DF}). Specifically, the Landauer principle, in contrast to a widespread opinion, is not the second law of thermodynamics nor is it equivalent to it, in fact it is a stricter bound. Clearly distinguishing the thermodynamic potentials (free energy $F$ and entropy $S$) from their non-equilibrium functional generalisations ($f$ and $s$ respectively), and the inequalities that characterise them,\footnote{It is worth remaking that  Eq. (\ref{eq:not-the-2nd-law}) is equivalent to the inequality $Q \leq T \Delta s$ (see \cite{Jarzynski99JSM96} for a generalised version thereof), which can be readily obtained from Eq. (\ref{eq:not-the-2nd-law}) by subtracting the quantity $\Delta E$ and using the relation  $Q=-(W-\Delta E)$, for the heat dumped into a thermal bath. Note how it superficially looks like the Clausius inequality $Q \leq T \Delta S$, which presents the thermodynamic potential $S$, i.e., the entropy, instead \cite{Fermi56Book}.} 
is a good practice that in the present case helps making sense of Landauer intuition that there is an extra unavoidable dissipation cost associated with information erasure. A similar warning regarding equilibrium vs. nonequilibrium quantities was raised as well in Ref. \cite{Hoerhammer08JSP133}.

\subsection{SSB adapted version of the Jarzynski equality}

We have illustrated above that the Jarzynski equality breaks in presence of spontaneous breaking of the fluctuation symmetry. However there is a natural and suggestive way to extend it to include such cases. As discussed in the standard literature, e.g., \cite{GoldenfeldBook}, in presence of SSB, the Gibbs distribution becomes inappropriate to describe the observed physics. 
However, just like the local thermal equilibrium  $\rho^{\text{eq},X}$ 
becomes the appropriate physical distribution, so the SSB adapted Jarzynski equality
\begin{align}
\langle e^{-\beta W}\rangle = e^{-\beta [F_X(t)-F(0)]}
\label{eq:J-SSB}
\end{align}
becomes physically appropriate, as our analysis showed. Here, $F=f[\rho^\text{eq}]$ denotes the standard ``global'' free energy, and $F_X=f[\rho^{\text{eq},X}]$ the ``local'' free energy associated to subspace $X$. In absence of SSB, it is $F_X=F$ and the standard formula is recovered. It is interesting to note how, in fact, Eq. (\ref{eq:J-SSB}) has been already employed to study Landauer's erasure \cite{Dillenschneider09PRL102}, while the fundamental reason behind its validity was not clarified.




\end{document}